\documentclass[pra,showpacs,twocolumn]{revtex4}
\usepackage{dcolumn}% Align table columns on decimal point
\usepackage{bm}% bold math
\usepackage{amssymb}
\usepackage{amsmath}
\usepackage{amsfonts}
\usepackage[]{graphicx}
\begin{document}
\bibliographystyle{prsty}
\title{Should the wave-function be a part of the quantum ontological state?
}
\author{Aur\'{e}lien Drezet}
\affiliation{Institut N\'eel UPR 2940, CNRS-University Joseph
Fourier, 25 rue des Martyrs, 38000 Grenoble, France} \pacs{}

\date{\today}

\begin{abstract}
We analyze the recent no go theorem by Pusey, Barrett and Rudolph
(PBR) concerning ontic and epistemic hidden variables. We define two
fundamental requirements for the validity of the result. We finally
compare the models satisfying the theorem with the historical hidden
variable approach proposed by de Broglie and Bohm.
\end{abstract}

\maketitle

\section{Introduction} \indent Recently, a new no
go theorem by M.~Pusey, J.~Barret and T.~Rudolph (PBR in the
following) was published~\cite{PBR}. The result concerns ontic
versus epistemic interpretations of quantum mechanics. Epistemic
means here knowledge by opposition to `ontic' or ontological  and is
connected with the statistical interpretation defended by Einstein.
This of course stirred much debates and discussions to define the
condition of validity of this fundamental theorem. Here, we discuss
two fundamental requirements necessary for the demonstration of the
result and also discuss the impact of the result on possible hidden
variable models. In particular, we will stress the difference
between the models satisfying the PBR theorem and those who
apparently contradict its generality.
\section{ The axioms of the PBR theorem}
\indent In order to identify the main assumptions and conclusions of
the PBR theorem we first briefly restate the original reasoning of
ref.~1 in a slightly different language. In the simplest version PBR
considered two non orthogonal pure quantum states
$|\Psi_1\rangle=|0\rangle$ and
$|\Psi_2\rangle=[|0\rangle+|1\rangle]/\sqrt{2}$ belonging to a
2-dimensional Hilbert space $\mathbb{E}$ with basis vectors
$\{|0\rangle,|1\rangle\}$. Using a specific (nonlocal) measurement
$M$ with basis $|\xi_i\rangle$ ($i\in[1,2,3,4]$) in
$\mathbb{E}\otimes\mathbb{E}$  (see their equation 1 in \cite{PBR})
they deduced that
$\langle\xi_1|\Psi_1\otimes\Psi_1\rangle=\langle\xi_2|\Psi_1\otimes\Psi_2\rangle=\langle\xi_3|\Psi_2\otimes\Psi_1\rangle=\langle\xi_4|\Psi_2\otimes\Psi_2\rangle=0$.
\indent In a second step they introduced hypothetical `Bell's like'
hidden variables $\lambda$ and wrote implicitly the probability of
occurrence $P_M(\xi_i;j,k)=|\langle\xi_i|\Psi_j\otimes\Psi_k
\rangle|^2$ in the form:
\begin{eqnarray}
P_M(\xi_i;j,k)=\int
P_M(\xi_i|\lambda,\lambda')\varrho_j(\lambda)\varrho_k(\lambda')d\lambda
d\lambda'
\end{eqnarray}
where $i\in[1,2,3,4]$ and $j,k\in[1,2]$. One of the fundamental
axiom used by PBR  (axiom 1) is an independence criterion at the
preparation which reads
$\varrho_{j,k}(\lambda,\lambda')=\varrho_j(\lambda)\varrho_k(\lambda')$.
In these equations we introduced the conditional `transition'
probabilities $P_M(\xi_i|\lambda,\lambda')$ for the outcomes $\xi_i$
supposing the hidden state $\lambda,\lambda'$ associated with the
two independent Q-bits are given. The fundamental point here is that
$P_M(\xi_i|\lambda,\lambda')$ is independent of $\Psi_{1},\Psi_{2}$.
This a very natural looking-like axiom (axiom 2) which was implicit
in ref.~1 and was not further discussed by the authors. We will see
later what are the consequence of its
abandonment.\\
\indent For now, from the definitions and axioms we obtain:
\begin{eqnarray}
\int P_M(\xi_1|\lambda,\lambda')\varrho_1(\lambda)\varrho_1(\lambda')d\lambda d\lambda'=0\nonumber\\
\int P_M(\xi_2|\lambda,\lambda')\varrho_1(\lambda)\varrho_2(\lambda')d\lambda d\lambda'=0\nonumber\\
\int P_M(\xi_3|\lambda,\lambda')\varrho_2(\lambda)\varrho_1(\lambda')d\lambda d\lambda'=0\nonumber\\
\int P_M(\xi_4|\lambda,\lambda')\varrho_2(\lambda)\varrho_2(\lambda')d\lambda d\lambda'=0.\nonumber\\
\end{eqnarray}
The first line implies $P_M(\xi_1|\lambda,\lambda')=0$ if
$\varrho_1(\lambda)\varrho_1(\lambda')\neq 0$. This condition is
always satisfied if $\lambda$ and $\lambda'$ are in the support of
$\varrho_1$ in the $\lambda$-space and $\lambda'$-space. Similarly,
the fourth line implies  $P_M(\xi_4|\lambda,\lambda')=0$ if
$\varrho_2(\lambda)\varrho_2(\lambda')\neq 0$ which is again always
satisfied if $\lambda$ and $\lambda'$ are in the support of
$\varrho_2$ in the $\lambda$-space and $\lambda'$-space. Finally,
the second and third lines imply  $P_M(\xi_2|\lambda,\lambda')=0$ if
$\varrho_1(\lambda)\varrho_2(\lambda')\neq 0$ and
$P_M(\xi_3|\lambda,\lambda')=0$ if
$\varrho_1(\lambda)\varrho_2(\lambda')\neq 0$.\\
\indent Taken separately these four conditions are not problematic.
But, in  order to be true simultaneously and then to have
\begin{eqnarray}
P_M(\xi_i|\lambda,\lambda')=0 \label{e6}
\end{eqnarray}
for a same pair of $\lambda,\lambda'$  (with $[i=1,2,3,4]$) the
conditions require that the supports of $\varrho_1$ and $\varrho_2$
intersect. If this is the case Eq.~3 will be true for any pair
$\lambda,\lambda'$ in the
intersection.\\
\indent However, this is impossible since from probability
conservation we must have
$\sum_{i=1}^{i=4}P_M(\xi_i|\lambda,\lambda')=1$ for every pair
$\lambda,\lambda'$. Therefore, we must necessarily have
\begin{eqnarray}\varrho_2(\lambda)\cdot\varrho_1(\lambda)=0 &
\forall \lambda\end{eqnarray} i.e. that $\varrho_1$ and $\varrho_2$
have nonintersecting supports in the $\lambda$-space. Indeed, it is
then obvious to see that Eq.~2 is satisfied if Eq.~4 is true. This
constitutes the PBR theorem for the particular case of independent
prepared states $\Psi_1,\Psi_2$ defined before. PBR generalized
their results for more arbitrary states using similar and astute
procedures described in ref.~1.\\
\indent If this theorem is true it would apparently make hidden
variables completely redundant since it would be always possible to
define a bijection or relation of equivalence between the $\lambda$
space and the Hilbert space: (loosely speaking we could in principle
make the correspondence $\lambda\Leftrightarrow\psi$). Therefore it
would be as if $\lambda$ is nothing but a new name
for $\Psi$ itself. This would justify the label `ontic' given to this kind of interpretation in opposition to `epistemic' interpretations ruled out by the PBR result.\\
\indent However, the PBR conclusion stated like that is too strong
as it can be shown by carefully examining the assumptions necessary
for the derivation of the theorem. Indeed, using the independence
criterion and the well known Bayes-Laplace formula for conditional
probability we deduce that the most general Bell's hidden variable
probability space should obey the following rule
\begin{eqnarray}
P_M(\xi_i;j,k)=\int P_M(\xi_i|\Psi_j,\Psi_k,\lambda,\lambda')\varrho_j(\lambda)\varrho_k(\lambda')d\lambda d\lambda'\nonumber\\
\end{eqnarray}
in which, in contrast to equation 1, the transition probabilities
$P_M(\xi_i|\Psi_j,\Psi_k,\lambda,\lambda')$ now depend explicitly on
the considered quantum states $\Psi_j,\Psi_k$. We point out that
unlike $\lambda$, $\Psi$ is in this more general approach not a
stochastic variable. This difference is particularly clear in the
ontological interpretation of ref. 3 where $\Psi$ plays the role of
a dynamic guiding wave for the stochastic motion of the particle.
Clearly, relaxing this PBR premise has a direct effect since we lose
the ingredient necessary for the demonstration of Eq.~4. (more
precisely we are no longer allowed to compare the product states
$|\Psi_j\otimes\Psi_k \rangle$ as it was done in ref.~1). Indeed, in
order for Eq.~2 to be simultaneously true for the four states
$\xi_i$ (where $P_M(\xi_i|\Psi_j,\Psi_k,\lambda,\lambda')$ now
replace $P_M(\xi_i|\lambda,\lambda')$) we must have
\begin{eqnarray}
P_M(\xi_1|\Psi_1,\Psi_1,\lambda,\lambda')=0,P_M(\xi_2|\Psi_1,\Psi_2,\lambda,\lambda')=0,\nonumber\\
P_M(\xi_3|\Psi_2,\Psi_1,\lambda,\lambda')=0,
P_M(\xi_4|\Psi_2,\Psi_2,\lambda,\lambda')=0.\nonumber\\
\end{eqnarray} Obviously, due to the explicit $\Psi$ dependencies, Eq.~6 doesn't anymore enter in conflict with
the conservation probability rule and therefore doesn't imply Eq.~4.
In other words the reasoning leading to PBR theorem doesn't run if
we abandon the axiom stating that
\begin{eqnarray}
P_M(\xi_i|\Psi_j,\Psi_k,\lambda,\lambda'):=P_M(\xi_i|\lambda,\lambda')
\end{eqnarray}
i.e. that the dynamic should be independent of $\Psi_{1},\Psi_{2}$.
This analysis clearly shows that Eq.~7 is a fundamental prerequisite
(as important as the independence criterion at the preparation) for
the validity of the PBR theorem~\cite{footnote1}. In our knowledge
this point was not yet discussed~\cite{footnote2}.
\section{Discussion}
\indent Therefore, the PBR deduction presented in ref.~1 is actually
limited to a very specific class of $\Psi$-epistemic
interpretations. It fits well with the XIX$^{th}$ like hidden
variable models using Liouville and Boltzmann approaches (i.e.
models where the transition probabilities are independent of $\Psi$)
but it is not in agreement with neo-classical interpretations, e.g.
the one proposed by de Broglie and Bohm~\cite{deBroglie}, in which
the transition probabilities $P_M(\xi|\lambda,\Psi)$ and the
trajectories depend explicitly and contextually on the quantum
states $\Psi$ (the de Broglie-Bohm theory being deterministic these
probabilities can only reach values 0 or 1 for discrete observables
$\xi$). As an illustration, in the de Broglie Bohm model for a
single particle the spatial position $\mathbf{x}$ plays the role of
$\lambda$. This model doesn't require the condition
$\varrho_1(\lambda)\cdot\varrho_2(\lambda)=|\langle\mathbf{x}|\Psi_1\rangle|^2\cdot|\langle\mathbf{x}|\Psi_1\rangle|^2=0$
for all $\lambda$ in clear contradiction with Eq.~4. We point out
that our reasoning doesn't contradict the PBR theorem \emph{per se}
since the central axiom associated with Eq.~7 is not true anymore
for the model considered. In other words, if we recognize the
importance of the second axiom discussed before (i.e. Eq.~7) the PBR
theorem becomes a general result which can be stated like that:\\
\indent i) If Eq.~7 applies then the deduction presented in ref.~1
shows that Eq.~4 results and therefore $\lambda\leftrightarrow\Psi$
which means that epistemic interpretation of $\Psi$ are equivalent
to ontic interpretations. This means that a XIX$^{th}$ like hidden
variable models is not really possible even if we accept Eq.~7 since
we don't have any freedom on the hidden variable density
$\rho(\lambda)$. \\
\indent ii) However, if Eq.~7 doesn't apply then the ontic state of
the wavefunction is already assumed - because it is a variable used
in the definition of $P_M(\xi|\lambda,\Psi)$. This shows that ontic
interpretation of $\Psi$ is necessary. This is exemplified in the de
Broglie-Bohm example: in this model, the "quantum potential" is
assumed to be a real physical field which depends on the magnitude
of the wavefunction, while the motion of the Bohm particle depends
on the wavefunction's phase. This means that the wavefunction has
ontological status in such a theory. This is consistent with the
central point of the PBR
paper but the authors didn't discussed that fundamental point.\\
\indent We also point out that in the de Broglie-Bohm ontological
approach the independence criterion at the preparation is respected
in the regime considered by PBR. As a concequence, it is not needed
to invoke
retrocausality to save epistemic approaches.\\
\indent  It is important to stress how Eq.~4, which is a consequence
of Eq.~7, contradicts the spirit of most hidden variable approaches.
Consider indeed, a wave packet which is split into two well
spatially localized waves $\Psi_1$ and $\Psi_2$ defined in two
isolated regions 1 and 2. Now, the experimentalist having access to
local measurements $\xi_1$ in region 1 can define probabilities
$|\langle\xi_1|\Psi_1\rangle|^2$. In agreement with de Broglie and
Bohm most proponents of hidden variables would now say that the
hidden variable $\lambda$ of the system actually present in box 1
should not depends of the overall phase existing between $\Psi_1$
and $\Psi_2$. In particular the density of hidden variables
$\varrho_\Psi(\lambda)$ in region $1$ should be the same for
$\Psi=\Psi_1+\Psi_2$ and $\Psi'=\Psi_1-\Psi_2$ since
$|\langle\xi_1|\Psi\rangle|^2=|\langle\xi_1|\Psi'\rangle|^2$ for
every local measurements $\xi_1$ in region 1. This is a weak form of
separability which is accepted even within the so exotic de Broglie
Bohm's approach but which is rejected for those models accepting
Eq.~4.\\
\indent This point can be stated differently. Considering the state
$\Psi=\Psi_1+\Psi_2$ previously discussed we can imagine a two-slits
like interference experiment in which the probability for detecting
outcomes $x_0$, ie., $|\langle x_0|\Psi\rangle|^2$ vanish for some
values $x_0$ while $|\langle x_0|\Psi_1\rangle|^2$ do not. For those
models satisfying Eq.~7 and forgetting one instant PBR theorem we
deduce that in the hypothetical common support of
$\varrho_{\Psi_1}(\lambda)$ and $\varrho_\Psi(\lambda)$ we must have
$P_M(\xi_0|\lambda)=0$ since this transition probability should
vanish in the support of $\Psi$. This allows us to present a
`\emph{poor-man}' version of the PBR's theorem: The support of
$\varrho_{\Psi_1}(\lambda)$ can not be completely included in the
support of $\varrho_\Psi(\lambda)$ since otherwise
$P_M(\xi_0|\lambda)=0$ would implies $|\langle
x_0|\Psi_1\rangle|^2=0$ in contradiction with the definition. PBR's
theorem is stronger than that  since it shows that in the limit of
validity of Eq.~7 the support of $\varrho_{\Psi_1}(\lambda)$ and
$\varrho_\Psi(\lambda)$ are necessarily disjoints. Consequently, for
those particular models the hidden variables involved in the
observation of the observable $\xi_0$ are not the same for the two
states $\Psi$ and $\Psi_1$. This is fundamentally different from de
Broglie-Bohm approach  where $\lambda$ (e.g. $\mathbf{x}(t_0)$) can
be the same for both states.\\
\indent This can lead to an interesting form of quantum correlation
even with one single particle. Indeed, following the well known
scheme of the Wheeler Gedanken experiment one is free at the last
moment to either observe the interference pattern (i.e. $|\langle
x_0|\Psi\rangle|^2=0$) or to block the path 2 and destroy the
interference (i.e. $|\langle x_0|\Psi_1\rangle|^2=1/2$). In the
model used by Bohm where $\Psi$ acts as a guiding or pilot wave this
is not surprising: blocking the path 2 induces a subsequent change
in the propagation of the pilot wave which in turn affects the
particle trajectories. Therefore, the trajectories will not be the
same in these two experiments and there is no paradox. However, in
the models considered by PBR there is no guiding wave since $\Psi$
serves only to label the non overlapping density functions of hidden
variable $\varrho_{\Psi_1}(\lambda)$ and $\varrho_{\Psi}(\lambda)$.
Since the beam block can be positioned after the particles leaved
the source the hidden variable are already predefined (i.e. they are
in the support of $\varrho_{\Psi}(\lambda)$). Therefore, the
trajectories are also predefined in those models and we apparently
reach a contradiction since we should have $P_M(\xi_0|\lambda)=0$
while we experimentally record particles with properties $\xi_0$.
The only way to solve the paradox is to suppose that some mysterious
quantum influence is sent from the beam blocker to the particle in
order to modify the path during the propagation and correlate it
with presence or absence of the beam blocker. However, this will be
just equivalent to the hypothesis of the de Broglie-Bohm guiding
wave and quantum potential and contradicts apparently the spirit and
the simplicity of $\Psi$-independent
models satisfying Eq.~7.\\
\section{An example}
\indent We point out that despite these apparent contradictions it
is easy to create an hidden variable model satisfying all the
requirements of PBR theorem. Let any state $|\Psi\rangle$ be defined
at time $t=0$ in the complete basis $|k\rangle$ of dimension $N$ as
$|\Psi\rangle=\sum_{k}^N\Psi_k|k\rangle$ with
$\Psi_k=\Psi_k'+i\Psi_k''$. We introduce two hidden variables
$\lambda$, and $\mu$ as the N dimensional real vectors
$\lambda:=[\lambda_1,\lambda_2..., \lambda_N]$ and
$\mu:=[\mu_1,\mu_2..., \mu_N]$. We thus write the probability
$P_M(\xi,t,\Psi)=|\langle\xi|U(t)\Psi\rangle|^2$ of observing the
outcome $\xi$ at time $t$ as
\begin{eqnarray}
\int P_M(\xi,t|\{\lambda_k,\mu_k\}_k)\prod_k^N\delta(\Psi_k'-\lambda_k)\delta(\Psi_k''-\mu_k)d\lambda_kd\mu_k\nonumber\\
=P_M(\xi,t|\{\Psi_k',\Psi_k''\}_k)=|\sum_k\langle\xi|U(t)|k\rangle\Psi_k|^2\label{ee}
\end{eqnarray}
where $U(t)$ is the Schrodinger evolution operator. Since $\Psi$ can
be arbitrary we thus generally have in this model
$P_M(\xi,t|\{\lambda_k,\mu_k\}_k)=|\sum_k\langle\xi|U(t)|k\rangle(\lambda_k+i\mu_k)|^2$.
The explicit time variation is associated with the unitary evolution
$U(t)$ which thus automatically includes contextual local or non
local influences (coming from the beam blocker for example). We
remark that this model is of course very formal and doesn't provide
a better understanding of the mechanism explaining the interaction
processes. The hidden variable model we proposed is actually based
on a earlier version shortly presented by Harrigan and Spekkens in
ref.~\cite{speckens}. We completed the model  by fixing the
evolution probabilities and by considering the complex nature of
wave function in the Dirac distribution. Furthermore, this model
doesn't yet satisfy the independence criterion if the quantum state
is defined as
$|\Psi\rangle_{12}=|\Psi\rangle_1\otimes|\Psi\rangle_2$ in the
Hilbert tensor product space. Indeed, the hidden variables
$\lambda_{12,k}$ and $\mu_{12,k}$ defined in Eq.~\ref{ee} are global
variables for the system 1,2. If we  write
\begin{eqnarray}
|\Psi\rangle_{12}=\sum_{n,p}^{N_1,N_2}\Psi_{12;n,p}|n\rangle_1\otimes|p\rangle_2\nonumber\\
 =\sum_{n,p}^{N_1,N_2}\Psi_{1;n}\Psi_{2;p}|n\rangle_1\otimes|p\rangle_2\label{eee}
\end{eqnarray}
the indices $k$ previously used become a doublet of indices $n,p$
and the probability
$P_M(\xi,t|\Psi_{12})=|\sum_{n,p}^{N_1,N_2}\langle\xi|U(t)|n,p\rangle_{12}\Psi_{12;n,p}|^2$
in Eq.~\ref{ee} reads now:
\begin{eqnarray}
\int P_M(\xi,t|\{\lambda_{12;n,p},\mu_{12;n,p}\}_{n,p}) \nonumber\\
\cdot
\prod_{n}^{N_1}\prod_{p}^{N_2}\delta(\Psi_{12;n,p}'-\lambda_{12;n,p})\nonumber\\ \cdot\delta(\Psi_{12;n,p}''-\mu_{12;n,p})d\lambda_{12;n,p}d\mu_{12;n,p}\nonumber\\
=P_M(\xi,t|\{\Psi_{12;n,p}',\Psi_{12;n,p}''\}_{n,p})
\end{eqnarray} which indeed doesn't show any explicit separation of the hidden variables density of states for subsystems 1 and 2. However, in the case where Eq.~\ref{eee} is valid
we can alternatively introduce new hidden variable vectors
$\lambda_1$, $\lambda_2$ and $\mu_1$, $\mu_2$ such that
$P_M(\xi,t|\Psi_{12})$ reads now:
\begin{eqnarray}
\int
P_M(\xi,t|\{\lambda_{1;n},\lambda_{2;p},\mu_{1;n},\mu_{2;n}\}_{n,p})
\nonumber\\ \cdot \prod_{n}^{N_1}\delta(\Psi_{1;n}'-\lambda_{1;n})\delta(\Psi_{1;n}''-\mu_{1;n})d\lambda_{1;n}d\mu_{1;n}\nonumber\\
\cdot \prod_{p}^{N_2}\delta(\Psi_{2;p}'-\lambda_{2;p})\delta(\Psi_{2;p}''-\mu_{2;p})d\lambda_{2;n}d\mu_{2;p}\nonumber\\
=P_M(\xi,t|\{\Psi_{1;n}',\Psi2'_{2;p},\Psi_{1;n}'',\Psi_{2;n}''\}_{n,p}).
\end{eqnarray}
Clearly here the density of probability
$\varrho_{12}(\lambda_1,\lambda_2,\mu_1,\mu_2)$ can be factorized as
$\varrho_{1}(\lambda_1,\mu_1)\cdot \varrho_{2}(\lambda_2,\mu_2)$
where
\begin{eqnarray}
\varrho_{1}(\lambda_1,\mu_1)=\prod_{n}^{N_1}\delta(\Psi_{1;n}'-\lambda_{1;n})\delta(\Psi_{1;n}''-\mu_{1;n})\nonumber\\
\varrho_{2}(\lambda_2,\mu_2)=\prod_{n}^{N_2}\delta(\Psi_{2;n}'-\lambda_{2;n})\delta(\Psi_{2;n}''-\mu_{2;n})
\end{eqnarray}
Therefore, the independence criterion at the preparation (i.e. axiom
1) is here fulfilled.\\
\indent Additionally, since by definition Eq.~8 and 10 are
equivalent we have
$P_M(\xi,t|\{\Psi_{1;n}',\Psi2'_{2;p},\Psi_{1;n}'',\Psi_{2;n}''\}_{n,p})=P_M(\xi,t|\{\Psi_{12;n,p}',\Psi_{12;n,p}''\}_{n,p})$.
Moreover, since $\Psi_{1;n}$ and $\Psi_{2;n}$ can have any complex
values the following relation holds for any value of the hidden
variables:
\begin{eqnarray}
P_M(\xi,t|\{\lambda_{1;n},\lambda_{2;p},\mu_{1;n},\mu_{2;n}\}_{n,p})\nonumber\\=P_M(\xi,t|\{\lambda_{12;n,p},\mu_{12;n,p}\}_{n,p})
\end{eqnarray} with
$\lambda_{12;n,p}+i\mu_{12;n,p}=(\lambda_{1;n}+i\mu_{1;n})(\lambda_{2;p}+i\mu_{2;p})$.
This clearly define a bijection or relation of equivalence  between
the hidden variables $[\lambda_{12}, \mu_{12}]$ on the one side and
$[\lambda_{1}, \mu_{1}, \lambda_{2}, \mu_{2}]$ on the second side.
Therefore, we showed that it is always possible to define hidden
variables satisfying the 2 PBR axioms: i) statistical independence
at the sources or preparation
$\varrho_{j,k}(\lambda,\lambda')=\varrho_j(\lambda)\varrho_k(\lambda')$
(if Eq.~9 is true) and ii) $\Psi$-indepedence at the dynamic level,
i.e., satisfying Eq.~7. We point out that the example discussed in
this section proves that the PBR theorem is not only formal since we
showed that it is indeed possible to build up explicitly  model
satisfying the two requirements of PBR theorem. This model is very
important since it demonstrate that the de Broglie Bohm approach is
not the only viable hidden theory. It is interesting to observe that
our model corresponds
to the case discussed in point i) of section 3 while Bohm's approach corresponds to the point labeled ii) in the same section.\\
\section{Conclusion}
\indent To conclude, we analyzed the PBR theorem and showed that
beside the important independence criterion already pointed out in
ref.~1 there is a second fundamental postulate  associated with
$\Psi$-independence at the dynamic level (that is our Eq.~7). We
showed that by abandoning this prerequisite the PBR conclusion
collapses. We also analyzed the nature of those models satisfying
Eq.~7 and showed that despite their classical motivations they also
possess counter intuitive features when compared for example to de
Broglie Bohm model. We finally constructed an explicit model
satisfying the PBR axioms. More studies would be be necessary to
understand the physical meaning of such hidden variable models.


\begin{thebibliography}{99}
\bibitem{PBR} M.~F.~Pusey, J.~Barrett, T.~Rudolph, Nature Phys. \textbf{8} (2012) 476.
\bibitem{speckens}
N.~Harrigan, R.~W.~Spekkens, Found.~Phys.\textbf{40} (2010) 125.
\bibitem{deBroglie} L.~de Broglie,  J. Phys. Radium\textbf{8}(1927) 225.
\bibitem{footnote1}
We point out that we will not save the naive interpretation of PBR's
theorem by using the `implicit' notation $\Lambda=(\lambda,\Psi)$
and by writing $\int P_M(\xi|\Lambda)\varrho(\Lambda)d\Lambda$ in
order to hide the $\Psi$ dependence
$P_M(\xi|\Lambda):=P_M(\xi|\lambda,\Psi)$. Indeed, at the end of the
day we have to compare different $\Psi$ states and the explicit
notation becomes necessary. The importance of Eq.~7 for PBR's result
can therefore not be avoided.
\bibitem{footnote2}
See also A.~Drezet, arXiv:1203.2475 (12 March 2012) for a detailed
discussion of PBR theorem.
\end{thebibliography}
\end{document}